\documentclass[aps,prl,preprint,showpacs]{revtex4}
\usepackage{bm}
\usepackage{graphicx}
\begin{document}
\title{Electron-phonon interaction via Pekar mechanism in nanostructures}
\author{B.~A.~Glavin}
\author{V.~A.~Kochelap}
\author{T.~L.~Linnik}
\affiliation{V.E.~Lashkarev Institute of Semiconductor Physics,
Ukrainian National
Academy of Sciences, Pr.~Nauki 41, Kiev 03028, Ukraine}
\author{K. W. Kim}
\affiliation{Department of Electrical and Computer Engineering,
North Carolina State University, Raleigh, NC 27695-7911, USA}
\begin{abstract}
We consider an electron-acoustic phonon coupling mechanism
associated with the dependence of crystal dielectric permittivity
on the strain (the so-called Pekar mechanism) in nanostructures
characterized by strong confining electric fields.  The efficiency
of Pekar coupling is a function of both the absolute value and the
spatial distribution of the electric field. It is demonstrated
that this mechanism exhibits a phonon wavevector dependence
similar to that of piezoelectricity and must be taken into account
for electron transport calculations in an extended field
distribution. In particular, we analyze the role of Pekar coupling
in energy relaxation in silicon inversion layers. Comparison with
the recent experimental results is provided to illustrate its
potential significance.
\end{abstract}


\maketitle

The electron-phonon interaction is one of the fundamental problems
in solid-state physics.  For coupling with acoustic phonons in
particular, much attention has been devoted to the two main
mechanisms in semiconductors: the deformation potential and
piezoelectric interaction. In the presence of an external electric
field $E$, however, an additional process appears associated with
the dependence of dielectric permittivity on the strain. Due to
this dependence, an acoustic phonon can induce an effective ac
electric field component and subsequently an additional coupling
with electrons. This mechanism was initially introduced by Pekar
\cite{Pekar1,Pekar2} for the problem of sound amplification by
drifting electrons. It was shown later \cite{Gulyaev} that the
Pekar mechanism of electron-phonon interaction is related directly
to the electrostriction effect. The induced ac electric field has
a piezo-like dependence on the phonon wavevector $q$ and can
roughly be characterized by an effective piezo-constant
\begin{equation}
\label{eq1prime}
\beta_{eff} \sim \varepsilon^2 E p,
\end{equation}
where $p$ is the photoelasticity constant and $\varepsilon$ is the
dielectric permittivity of the unstrained crystal. In ordinary
bulk crystals with $E\alt 10^4~V/cm$, $\beta_{eff}$ is quite
small. As a result, Pekar initially concentrated on the materials
with a very large $\varepsilon$ (e.g., $ \varepsilon \sim 2000$
for ${\rm BaTiO_3}$) and electrostriction effect. Efficient
acousto-electric coupling for such materials was observed
experimentally in the seventies \cite{Kucherov}.

As the dimension of the sample structure shrinks, interaction
mechanisms that are not allowed in bulk materials can manifest
themselves. For instance, modulation of the quantum well width and
effective mass by strain gives rise to the so-called macroscopic
deformation potential \cite{macdp}. Similarly, the "traditional"
mechanisms can exhibit different features. In this paper, we
demonstrate that the Pekar mechanism is essential for
nanostructures with strong confining electric fields
\cite{Ando,nitride}. The interaction efficiency is dependent on
the spatial scale of the electric field localization and the
phonon wavelength. As a specific example, we consider
low-temperature electron energy relaxation in n-type $(100)$ Si
inversion layers. Our calculation shows that due to the Pekar
mechanism, the dissipated power in these non-piezoelectric
materials can exhibit a piezo-like $T^3$ dependence. This finding
provides a clear explanation for a recent experimental observation
of similar dependence \cite{Pudalov}.

Let us start with a general analysis of the Pekar mechanism for
the case of a nonuniform electric field distribution. The
variation of dielectric permittivity under strain $u_{ij}$ can be
written in the form \cite{Tucker}:
\begin{equation}
\label{eq:n2}
\delta\varepsilon_{ij} = \varepsilon_{ij}^2 p_{iklm} u_{lm},
\end{equation}
where $p_{iklm}$ is the photoelasticity tensor (we could use the
electrostriction tensor instead; however, the use of
photoelasticity tensor allows one to extract explicitly the
dependence of electrostriction on $\varepsilon$, which follows
from the Clausius--Mossotti model \cite{AshMer}). The form of
photoelasticity tensor is determined by the symmetry of the
crystal. In the diamond and zinc-blende structures, there are only
three independent nonzero components of this tensor denoted
usually as $p_{11}$, $p_{12}$, and $p_{44}$ \cite{Tucker}.
Moreover, $\varepsilon_{ij} = \delta_{ij} \varepsilon$.

The electrostriction effect in a static external field $\bm E$
leads to appearance of an effective field $\tilde{\bm{E}}=-\nabla
{\tilde \phi}$. The equation for potential $\tilde \phi$ can be
derived from the equation for the electric displacement $\bm D$:
\begin{equation}
\label{eq:n3}
\nabla\cdot {\bm D} = 0,~D_i = \varepsilon_{ij} E_k^{(\Sigma)},
\end{equation}
where ${\bm E}^{(\Sigma)}={\bm E}+\tilde {\bm E}$ is the total
electric field in the crystal. Assuming that the strain is small,
we obtain
\begin{equation}
\label{eq:n5}
 \nabla^2 {\tilde \phi} =\frac{1}{\varepsilon}
\frac {\partial}{\partial x_i} (\delta\varepsilon_{ik} E_{k}).
\end{equation}
In the simplest situation, the external electric field is ${\bm E}
=(0,0,E(z))$, which corresponds to the confining field in a
quantum well. We also assume an elastically uniform medium with
the functional dependence of $u_{ij},\delta\varepsilon_{ij} \sim
\exp [i (-\omega t + q_z z + {\bm q}_{||} {\bm \varrho})]$ and
${\tilde \phi}={\tilde \phi}(z)\exp [i(-\omega t +{\bm q}_{||}
{\bm \varrho})]$, where ${\bm q}_{||} = (q_x,q_y)$ and ${\bm
\varrho} = (x,y)$ is the coordinate vector in the quantum well
plane. Under these conditions, Eq.~(\ref{eq:n5}) simplifies as
\begin{equation}
\label{eq:n6}
\frac{d^2{\tilde \phi}}{dz^2} -q^2_{||}{\tilde \phi} =
G(z),~~G(z) \equiv \left( E(z)
 {\frac {\partial \delta\varepsilon_{iz}} {\partial x_i}} +
\delta\varepsilon_{zz}
 {\frac {d E(z)} {d z}} \right) \frac{1}{\varepsilon},
\end{equation}
resulting in
\begin{equation}
\label{eq:n7}
 {\tilde \phi}=-{\frac 1 {2 q_{\parallel}}} e^{q_{\parallel}z} \int^{+\infty}_z
 G(z')e^{-q_{\parallel}z'} dz'-{\frac 1 {2 q_{\parallel}}} e^{-q_{\parallel}z} \int^z_{-\infty}
 G(z')e^{q_{\parallel}z'} dz'.
\end{equation}
It is important to note two important features of $\tilde \phi$.
First, $\tilde \phi$ is not a plane wave as a function of $z$,
which is a direct result of the assumed nonuniform character of
the electric field. Second, $\tilde \phi$ strongly depends on the
spatial domain of electric field localization, $d$. For $q_z d \gg
1$ and $ q_{||}d \gg 1$, the estimate of Eq.~(\ref{eq1prime})
applies. For $q_z d\ll 1$ and $ q_{||}d \ll 1$, however, the
induced potential is suppressed substantially with a factor $q d$.

Let us consider a specific example, namely, the process of energy
relaxation in n-type $(100)$ Si inversion layers at low
temperatures. The built-in confining electric fields in such
structures can be as high as several $MV/cm$. In this case, the
role of Pekar mechanism is particularly important since silicon
itself is not a piezoelectric material and the Pekar contribution
is expected to be dominant at low temperatures, where the
deformation potential interaction is less effective. Of course,
the Pekar mechanism also provides a contribution to the momentum
relaxation rate that determines the mobility. However, at low
temperatures the momentum relaxes mainly due to the elastic
scattering by various imperfections, and the contribution of
phonons is hardly measurable. Thus, we concentrate on energy
relaxation.

The electron potential well near the Si/${\rm SiO_2}$ interface
and the electric field associated with the inversion layer are
shown schematically in Fig.~1.  The strong static electric field
$E=-d \phi(z)/ d z $ associated with the inversion layer confines
the electrons in a thin silicon layer near the interface. The
electrostatic potential $\phi(z)$, quantized electron levels and
wavefunctions $\psi({\bm r})=\chi(z) \exp (i {\bm k}{\bm
\varrho})$ are determined from the self-consistent solution of
Poisson and Schr\"{o}dinger equations \cite{Ando}. In the
following, we assume that only the ground electron subband is
populated at low temperatures.

The induced potential is determined by Eq.~(\ref{eq:n6}). The
boundary conditions at the Si/${\rm SiO_2}$ interface ($z=0$)
under a small strain are:
\begin{equation}
\label{eq:n7a}
  {\tilde \phi^{(s)}} ={\tilde \phi^{(ins)}},~~
\varepsilon^{(s)} {\tilde E^{(s)}_z} +
\delta \varepsilon_{zz}^{(s)} E^{(s)}=
 \varepsilon^{(ins)} {\tilde E^{(ins)}_z} + \delta \varepsilon_{zz}^{(ins)}
 E^{(ins)},
\end{equation}
where ${\tilde E_z} = -d {\tilde \phi} / dz$. For the considered
geometry, the following photoelasticity terms are relevant:
\begin{eqnarray}
\label{eq:n9}
\delta \varepsilon_{xz}=2{\varepsilon^2} p_{44} u_{xz}, \nonumber\\
\delta \varepsilon_{yz}=2{\varepsilon^2} p_{44} u_{yz},\\
\delta \varepsilon_{zz}={\varepsilon^2} (p_{11} u_{zz}+p_{12}
(u_{yy}+u_{xx})), \nonumber
\end{eqnarray}
where $x,y,z$ are the symmetry axis of the crystal. For
simplicity, we disregard the mismatch of elastic and
photoelasticity constants in Si and ${\rm SiO_2}$ layers. We also
assume that the actual phonon wavelength exceeds the thickness of
the depletion layer in silicon (see Fig.~1) and yet is much less
than the thickness of the ${\rm SiO_2}$ layer. The former
assumption can be justified at low enough temperatures, where the
contribution of the Pekar mechanism exceeds that of the
deformation potential. In case when the latter condition is
violated, the induced potential becomes weak in accordance with
the general analysis given above. In fact, this restricts the
parameters of actual structures where the Pekar mechanism is
important.

Under these assumptions, $\int dz \chi^2 (z) {\tilde \phi}(z)
\approx {\tilde \phi}(0)$, leading to
\begin{equation}
\label{eq:n14} \tilde{\phi} (0) =\frac{\varepsilon^{(s)}
\varepsilon^{(ins)}}{\varepsilon^{(s)} +\varepsilon^{(ins)}}
\frac{E_s}{q_{||}+iq_z} \left( (p_{44}q_{||}+iq_z p_{11})u_z
+\left(p_{44}\frac{q_z}{q_{||}} +ip_{12}\right)({\bm u}_s {\bm
q_{||}})\right).
\end{equation}
Here, $E_s$ and ${\bm u}_s$ are the electric field and the phonon
displacement at the silicon side of the interface, respectively.

Using Eq.~(\ref{eq:n14}), we calculate the electron scattering
rates $W^{q_z;\pm}_{kk'}$ for electron transition ${\bm
k}\rightarrow {\bm k'}$ with absorption ($+$) or emission ($-$) of
an acoustic phonon:
\begin{equation}
\label{eq:n15}
W^{q_z;\pm}_{{\bm kk}'}={\frac {\pi e^2 (E_s)^2}{V\rho \omega_q}}\left(
{\frac {\varepsilon^{(s)} \varepsilon^{(ins)}}{\varepsilon^{(s)}
+\varepsilon^{(ins)}}}\right)^2
\mid \alpha_l(\theta)\mid ^2 \delta_{{\bm k'},{\bm k}\mp {\bm q_{||}}}
\delta(\epsilon_k-\epsilon_{k'}\mp \hbar \omega_q)
\left\{ \begin{array}{l} N_q+1\\N_q
\end{array} \right.
\end{equation}
where $\epsilon_k$ is the electron energy and $N_q$ represents the
Planck population of the phonon modes characterized by the lattice
temperature $T$. The index $l$ of the function $\mid
\alpha_l(\theta)\mid$ denotes the type of the phonon modes. In
particular,
\begin{equation}
\label{eq:n16}
\mid \alpha_{LA}(\theta)\mid^2=\left(p^2_{11}+(4p^2_{44}-2p^2_{11}+2p_{11}p_{12})\sin^2\theta+
(p^2_{11}+p^2_{12}-2p_{11}p_{12}-4p^2_{44})\sin^4\theta\right),
\end{equation}
\begin{equation}
\label{eq:n17}
\mid \alpha_{TA}(\theta)\mid^2=\left(p^2_{44}(\sin^2\theta-\cos^2\theta)^2+
(p_{11}-p_{12})^2 \sin^2\theta \cos^2\theta\right),
\end{equation}
where $\theta$ is the angle between the phonon wave vector $\bm q$
and the $z$ axis, $V$ and $\rho$ are the normalizing volume and
density, respectively, and $\omega_q=s_{l,t} q$ is the phonon
frequency. $\alpha_{TA}$ corresponds to the contribution of the
transverse phonons with a vertical polarization ($\bm u$ lies in
the plane formed by the $z$ axis and $\bm q$); the contribution of
the phonons with a horizontal polarization ($\bm u$ is parallel to
the interface) is zero.
The power $Q$ dissipated by electrons is given as
\cite{gantmakher}:
\begin{equation}
\label{eq:n20}
Q=g S^{-1} \Sigma_{{\bm k,k}',q_z}(\epsilon_k-\epsilon_{k'})
(W_{{\bm kk}'}^{q_z;+} + W_{{\bm kk}'}^{(q_z;-)}) f_{T_e}({\bm k})
\left(1-f_{T_e}({\bm k}')\right),
\end{equation}
where $g$ is the electron degeneracy ($g=4$ in a $(100)$ inversion
layer taking into account both the spin and valley degeneracy) and
$S$ is the cross sectional area for normalization. In the
expression for $Q$, we assume that the electron-electron
interaction establishes a Fermi distribution for electrons with an
electron temperature $T_e$.

At a low temperature, the small-angle (Bloch-Gr\"{u}nizen)
scattering regime is realized. For degenerate electrons after the
standard transformations, we obtain a piezo-like temperature
dependence:
\begin{eqnarray}
\label{eq:n21}
Q=C(T^3_e-T^3), \nonumber \\
C=(C_{LA}+C_{TA}){\frac 1 {\pi^{7/2}}}
{\frac {m^2 (E_s)^2}{\hbar^5 n_e^{1/2}}}\frac {e^2}{\rho} \zeta(3)
\left({\frac {\varepsilon^{(s)} \varepsilon^{(ins)}}{\varepsilon^{(s)}+\varepsilon^{(ins)}}}\right)^2,\\
C_{LA}={\frac 1 {s^2_{l}}}{\frac {\pi} 8} \left( 4 p^2_{44}+2 p_{11}p_{12}+3 p^2_{11}+
3p^2_{12}\right),\nonumber\\
C_{TA}={\frac 1 {s^2_{t}}}{\frac {\pi} 8} \left( 4 p^2_{44}+(p_{11}-p_{12})^2\right),\nonumber
\end{eqnarray}
where $m$ is the electron effective mass for the in-plane motion,
$n_e$ is the electron concentration in the inversion layer, and
$\zeta$ is the Riemann zeta-function, $\zeta(3)\approx 1.20$.

For a numerical estimation, we use the following material
parameters: $\varepsilon_{Si}=12$, $\rho=2~g/cm^3$,
$\varepsilon_{\rm SiO_2}=4$, $s_{l}=9 \times 10^5~cm/s$,
$s_{t}=5.4\times 10^5~cm/s$, and $m=0.19 m_0$ [for the (100)
interface]. The values of the photoelasticity tensor are
 $p_{11}=-0.093$, $p_{12}=0.026$, and $p_{44}=-0.05$ \cite{Levine}.
As mentioned previously, the electron density and electric field
at the interface must be determined by a self-consistent
procedure. As a rough estimate, we adopt $E_s = 3\times 10^5~V/cm$
and $n_e=5\times 10^{11}~cm^{-2}$, which provide $C=10^{-4}~W/K^3
m^2$. This corresponds to the energy relaxation time of about $
1~ms$ at the electron temperature of $T_e \sim 1$~K and $T\sim
T_e$. 

In a recent experiment on silicon metal-oxide-semiconductor
field-effect transistors \cite{Pudalov}, the cubic dependence of
$Q$ on $T_e$ was observed at low temperatures, which was followed
by a $T_e^5$ dependence for $T_e > 0.6 $~K characteristic for the
deformation potential coupling.  The authors of
Ref.~\onlinecite{Pudalov} attributed the $T_e^3$ dependence to the
appearance of effective piezoelectric properties due to the
specific structure of the interface. Indeed, the interface reduces
the symmetry of the system which can give rise to nonzero
piezoelectricity in the vicinity of interface. However, for the
phonon wavelength exceeding the thickness of this "piezoelectric"
layer, the  piezoelectric potential induced by the phonon is
suppressed (this is similar to the case of the Pekar potential for
narrow regions of electric field localization). Therefore, the
observed cubic dependence is more likely to be due to the Pekar
mechanism.
Note also, that the obtained above value of $C$ is close to that measured
experimentally \cite{Pudalov}.

In summary, we show that the Pekar mechanism of electron-phonon
interaction (related to the electrostriction effect) can become
important in nanostructures due to the presence of strong
confining electric fields and must be considered along with the
deformation potential and piezoelectric mechanisms. The
effectiveness of Pekar coupling depends on both the absolute value
and the spatial distribution of the electric field. An estimation
of power dissipation by this mechanism in a silicon inversion
layer is in good agreement with a recent experiment.

We are indebted to V.N. Piskovoi for many valuable comments and
discussions. The work was supported in part by AFOSR and CRDF
(grant UE2-2439-KV-02).

\begin{figure}
\caption{ (a) Schematic illustration of the potential well in a
n-type Si inversion layer. The thickness of the depletion layer is
denoted by $d_{depl}$. (b) $z$-dependence of the confining
electric field in the inversion layer. }
\end{figure}

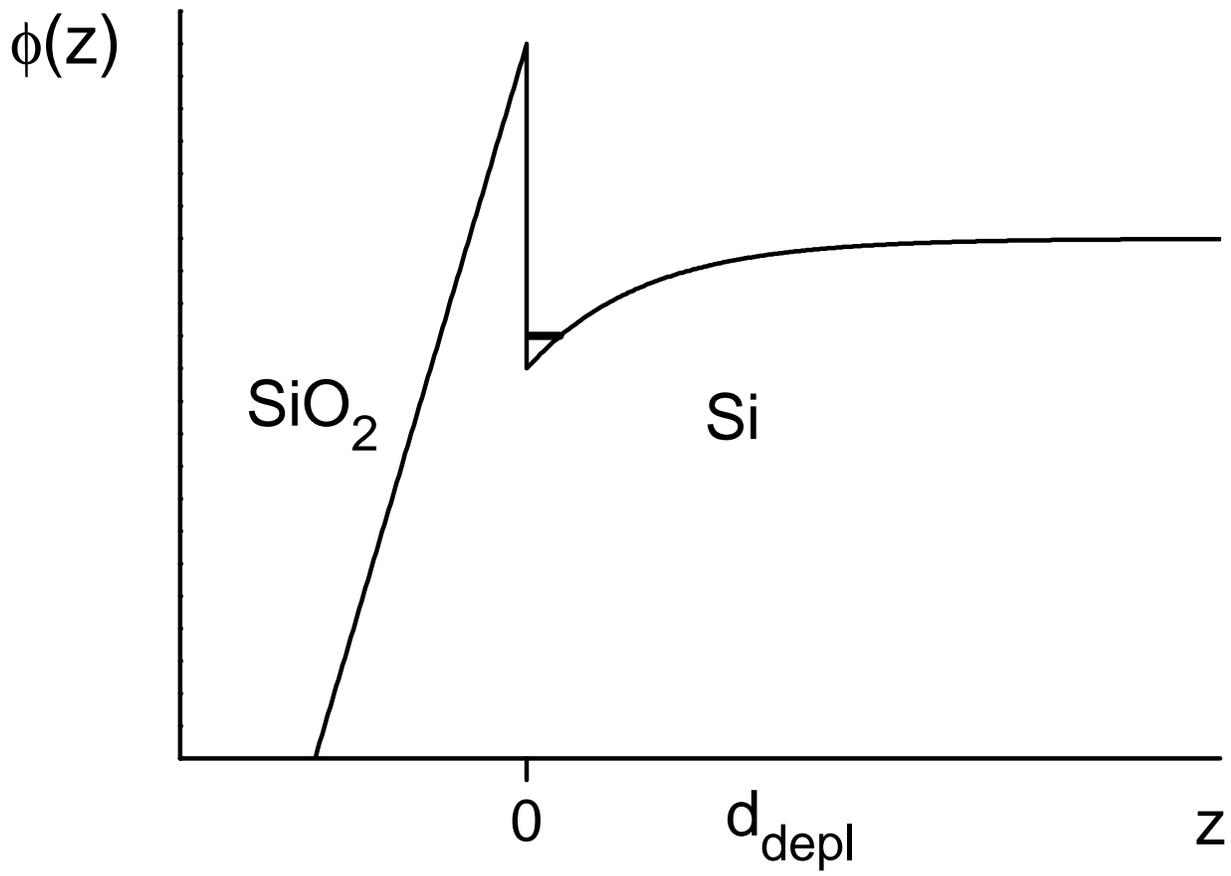
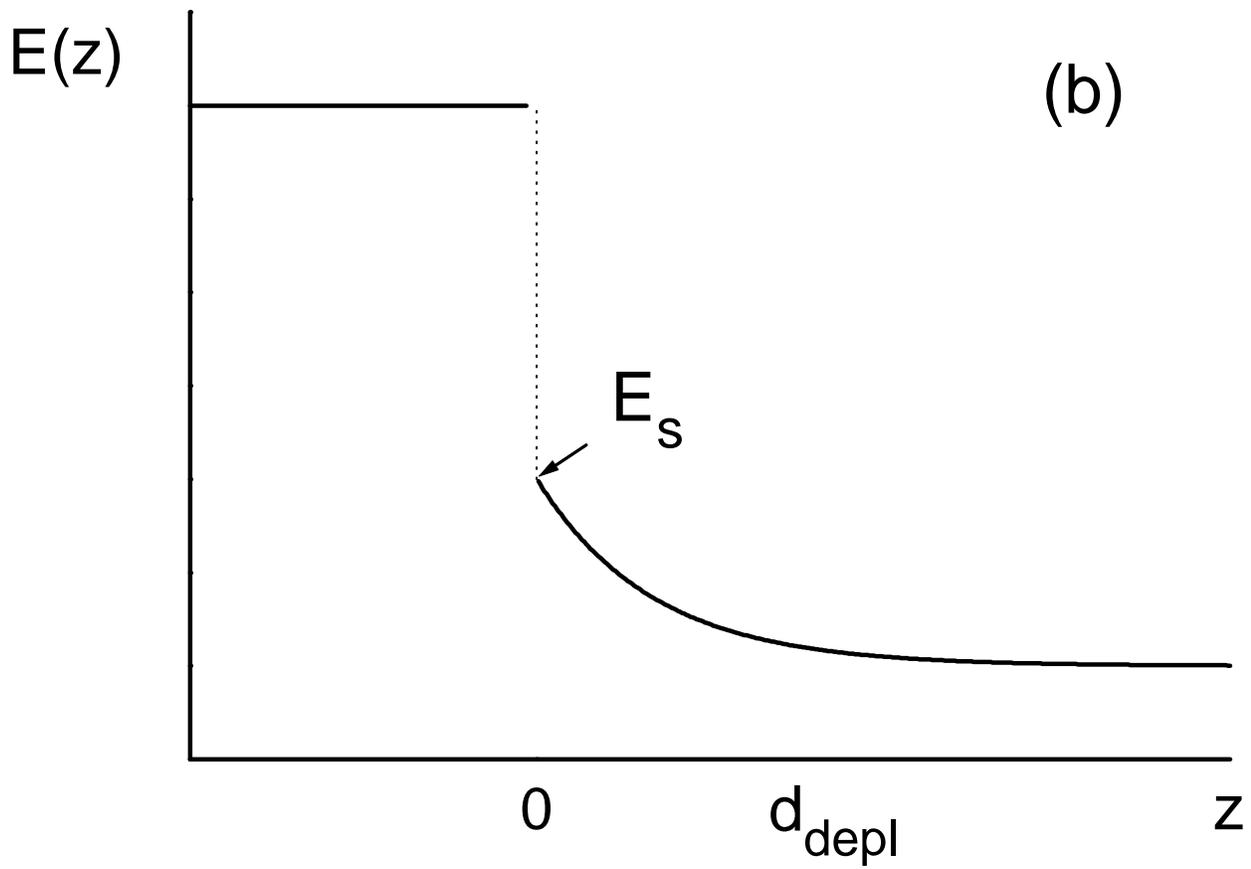
\end{document}